\documentstyle[11pt,epsfig]{article}
\newcommand{\beq}{\begin{equation}}
\newcommand{\eeq}{\end{equation}}
\newcommand{\beqa}{\begin{eqnarray}}
\newcommand{\eeqa}{\end{eqnarray}}

\begin{document}

\hfill FZJ-IKP(TH)-1998-04

\hfill nucl-th/9804005

\vspace{1in}

\begin{center}

{{\Large\bf Low--momentum effective theory for nucleons}}

\end{center}

\vspace{.3in}
\begin{center}
{\large 
E.~Epelbaoum,$^{a,b}$\footnote{email: evgeni.epelbaum@hadron.tp2.ruhr-uni-bochum.de}  
W.~Gl\"ockle,$^a$\footnote{email: walter.gloeckle@hadron.tp2.ruhr-uni-bochum.de} 
Ulf-G. Mei{\ss}ner$^b$\footnote{email: Ulf-G.Meissner@fz-juelich.de}}

\bigskip

$^a${\it Ruhr-Universit\"at Bochum, Institut f{\"u}r
  Theoretische Physik II\\ D-44870 Bochum, Germany} 

\bigskip

$^b${\it Forschungszentrum J\"ulich, Institut f\"ur Kernphysik 
(Theorie)\\ D-52425 J\"ulich, Germany}

\end{center}

\vspace{.9in}
\thispagestyle{empty} 

\begin{abstract}\noindent
Starting from a precise two--nucleon potential,
we use the method of unitary transformations to construct an
effective potential that involves only momenta less than
a given maximal value. We describe this method for an S--wave
potential of the Malfliet--Tjon type. It is demonstrated that
the bound and scattering state spectrum calculated within the 
effective theory agrees exactly with the one based on the original
potential. This might open an avenue for the construction of
effective chiral few--nucleon forces and for a systematic treatment
of relativistic effects in few--body systems. 
\end{abstract}

\vfill

\pagebreak


\noindent {\bf 1.} Chiral perturbation theory for two and more nucleons
became a subject of a great research interest in the past few 
years, see e.g. the pioneering work in~\cite{weinberg,bira}. 
One hopes to be able to clarify the structure 
of nuclear forces in this way. However, only the low--momentum
matrix elements of nuclear forces may be systematically treated 
in this approach since it is based on a consistent power counting
of small momenta and pion masses compared to the typical hadronic
scale of $\Lambda_{\rm had} \simeq$~1~GeV (for some recent work along
these lines see e.g. refs.~\cite{kswold,EGM}). A natural problem arises
due to the appearance of shallow nuclear bound states indicating
a breakdown of perturbation theory. Furthermore,
solving the Lippmann--Schwinger equation for constructing the deuteron
necessarily involves momenta $|\vec{p}\, | > \Lambda_{\rm had}$. For
such momenta, the chiral effective potential constructed according to
the conventional power counting rules is no longer applicable. This
is witnessed by the fact that in the calculations for the two--nucleon
system in~\cite{bira}, an additional ad hoc cut--off to tame the
high--momentum components had to be introduced. To be more precise, 
this cut--off function is not commensurate with the underlying chiral power
counting since it introduces an infinite string of local operators with 
increasing dimension. One might therefore
question the validity or usefulness of such an approach alltogether.
For recent discussions of this subject see~\cite{silas} and a
different power counting  scheme has been presented in~\cite{ksw}. 
A similar problem arises in standard few-- and many--body 
calculations based on realistic
nucleon--nucleon (NN) forces. The potentials, if derived from
meson--exchange diagrams, are generally based on a
non--relativistic expansion in powers of momenta over the nucleon
mass and are then used in various types of bound state
equations. These usually involve integrations over a much larger 
range of momenta as used in the construction of the NN potentials.
The same is of course also true for the various phenomenological NN
forces, which are chosen more or less ad hoc (with the exception of
the pion tail).
This affects in a non--trivial way the calculation of observables,
such as masses, levels or electromagnetic response functions. For a
recent discussion, see e.g. ref.\cite{kagl}.
In this general context the question of
the existence and the  properties of a low--momentum effective theory 
for nucleons are thus of great importance. We show here that it is indeed
possible to construct an effective two--nucleon potential 
from a given realistic potential which
involves only low momenta, i.e. momenta below a chosen momentum cut--off,
but which gives exactly the same results for bound and scattering states.
The cut--off scale introduced in our approach should be considered a physical
quantity since it defines the Hilbert space in which the theory operates.
This is different from the cut--off in a form factor or vertex function.
It is important to stress that our approach differs from the treatment
of the Schr\"odinger equation in an effective field theory framework proposed
by Lepage~\cite{lepage}. In his approach, an effective field theory for nucleons
only is constructed for very low momenta and eventually pions are added. While
that is certainly a valid framework, we intend to stay closer to the already
existing nuclear physics knowledge in that our approach will eventually allow
to match the low--momentum theory in a well--defined way to the highly successful
meson--exchange pictures of the nucleon--nucleon force. 
The results presented here should therefore be considered as a first step in 
a bigger program. Finally, we remark that while it seems to be known that such
an exact momentum space projection can be done, to our knowledge this program
has never been carried out before.

\medskip

\noindent {\bf 2.} To be specific, we consider a momentum--space Hamiltonian for the
two--nucleon system of the form
\begin{equation}
{\cal H} (\vec{p},\vec{p}\, '  ) = {\cal H}_0 (\vec{p}\, ) \, 
\delta (\vec{p} - \vec{p}\,'  ) 
+ {\cal V}  (\vec{p},\vec{p}\, ' ) \,\, ,
\end{equation} 
where ${\cal H}_0$ stands for the kinetic energy and the explicit
form of the NN potential will be specified later. 
For illustrative purposes we stick here to a simple S--wave potential. 
Note, however, that the inclusion of spin and isospin
dependent potentials can be handled along the lines outlined here.
Our aim is to  decouple the low and high momentum
components of this two--nucleon potential
using the method of unitary transformation~\cite{okubo,FST}.
For achieving that, we introduce the projection operators
\begin{eqnarray}
\eta    &=& \int d^3 p \, 
| \vec{p} \,\rangle \langle {\vec p}\, | \,\, , \quad 
| \vec{p}\,| \le \Lambda \,\, ,   
\nonumber \\
\lambda &=& \int d^3 p \,
| \vec{p} \,\rangle \langle {\vec p}\, | \,\, , \quad 
| \vec{p}\,| > \Lambda \,\, , 
\end{eqnarray}
where $\Lambda$ is a momentum cut--off whose value will be specified later and
$\eta$ ($\lambda$) is a projection operator onto low (high) momentum states
with $\eta^2 = \eta$, $\lambda^2 = \lambda$, $\eta \lambda
= \lambda \eta =0$ and $\lambda + \eta = {\bf 1}$. To be precise, 
the separation into low and high momentum components is to be
understood in a limiting sense, we always consider $\lim_{\epsilon\to
  0} (\Lambda - \epsilon)$. 
In this basis, the
Schr\"odinger equation takes the form
\begin{equation}
\label{Seq}
\left( \begin{array}{cc} \eta {\cal H} \eta & \eta {\cal H} \lambda \\ 
\lambda {\cal H} \eta & \lambda  {\cal H}
\lambda \end{array} \right) \left( \begin{array}{c}  \eta \Psi \\ 
\lambda \Psi \end{array} \right)
= E  \left( \begin{array}{c} \eta \Psi  \\ 
\lambda \Psi  \end{array} \right)
\quad .
\end{equation}
We now perform a unitary transformation  of the type
\begin{equation}
{\cal H} \to {\cal H}' = U^\dagger \, {\cal H} \, U \,\, \, ,
\end{equation}
so that $\eta {\cal H}' \lambda = \lambda {\cal H}' \eta = 0$.
The corresponding unitary operator $U$ is parametrized
in terms of an operator $A$, following Okubo~\cite{okubo}:
\begin{equation}
\label{A}
U = \left( \begin{array}{cc} (1 + A^\dagger A )^{- 1/2} & - 
A^\dagger ( 1 + A A^\dagger )^{- 1/2} \\
A ( 1 + A^\dagger A )^{- 1/2} & (1 + A A^\dagger )^{- 1/2} \end{array} \right)
\end{equation}
and $A$ satisfies the condition $A = \lambda A \eta$.
The requirement of decoupling
the two spaces leads to the following nonlinear integral equation
\begin{equation}\label{Aeq}
\lambda \left( {\cal H} - [ A, \, {\cal H} ] - A {\cal H} A \right) \eta = 0
\end{equation}
for the operator $A$. 
In the context of the nuclear many--body theory one often introduces a
mean field single particle basis, which defines a complete set of
N--particle states. A low--energy subgroup of states form a model space
and one is interested in effective interactions acting in that model space
such that the same low energy spectrum results as for the underlying 
N--body Hamiltonian. A way to arrive at that effective interaction is to 
decouple by a suitable transformation the two spaces (model space and 
the rest space), which leads to a decoupling equation of exactly the
form Eq.(\ref{Aeq}). In that context it is often reformulated into a 
linear form on a two--body  level using the exactly known interacting 
two--body states (some  references are e.g.~\cite{kuo}\cite{SO}).
This is indeed a feasible way to proceed also
in our context, as will be shown in a forthcoming
article. Here, however, we solve directly the nonlinear equation~(\ref{Aeq}).
If we denote by $\vec{q} \, (\vec{p}\, )$ a momentum from the
$\eta \, (\lambda)$--space, Eq.(\ref{Aeq}) takes the form\footnote{Our
notation is such that $Q(\vec{p},\vec{q})$ stands for the corresponding matrix
element $\langle \vec{p} \, | Q | \vec{q} \, \rangle$ for any operator $Q$.} 

\begin{eqnarray}\label{eq:V}
{\cal V} (\vec{p} ,\vec{q}\, ) &-& \int  d^3q' \, 
A( \vec{p}, \vec{q}\, ' ) {\cal V}( \vec{q}\,' , \vec{q}\,)  + \int  d^3p' \, 
{\cal V} (\vec{p} , \vec{p}\,' ) A (\vec{p}\,' , \vec{q}\, )
  \nonumber\\
&-& \int  d^3q' \,  d^3p' \, 
A(\vec{p} , \vec{q}\,' ) {\cal V} (\vec{q}\, ' , \vec{p}\, ' ) 
A(\vec{p}\, ' , \vec{q}\, )
= (E_{\vec{q}} - E_{\vec{p}}) \, A(\vec{p} , \vec{q} \,) \quad.
\end{eqnarray}
The quantities $E_{\vec{q} \, , \vec{p}\,}$ are the kinetic energies
related to the corresponding three--momenta.
This equation can only be solved numerically. This is most easily done
by iteration starting with
\beq\label{eq:start}
A = \frac{{\cal V}(\vec{p}, \vec{q}\,)}{E_{\vec{q}} - E_{\vec{p}}}~.
\eeq 
After four iterations, we then perform
an average over the values of the operator $A$ with different weight
factors. This allows to speed up the convergence considerably (details
on this procedure will be published elsewhere).
We also provide a regularization  scheme for the singularities of the
operator $A$, which  arise by solving this equation, i.e. at the
cut--off momentum (as becomes obvious from
Eq.(\ref{eq:start})).\footnote{In contrast, the effective potential is
  well--behaved.}
To be specific, we redefine the original potential ${\cal V}(\vec{p} ,
\vec{p}\,')$ by multiplying it with some smooth functions
$f(\vec{p}\,)$ and $f(\vec{p}\,')$ which are zero in some
neighborhood of the point $|\vec{p}\, | = \Lambda$ and one elsewhere.
The precise form of this procedure is of no interest for the following
and will thus not be discussed in detail here. We only add that the
regularization is chosen mild enough that it has no effect on the
observables, as will be illustrated later on for a specific example.

\medskip

\noindent {\bf 3.} We now restrict ourselves to the NN S--waves. To be specific, 
consider a momentum--space Malfliet--Tjon~\cite{MT} potential with an attractive
and a repulsive part
\begin{equation}\label{PotMT}
{\cal V}_{\rm MT} (\vec{q}_1, \vec{q}_2) = \frac{1}{2\pi^2} \biggl( 
\frac{V_R}{t + \mu_R^2} -
\frac{V_A}{t + \mu_A^2} \biggr)~, 
\end{equation}
with $t = (\vec{q}_1 - \vec{q}_2)^2$. We choose
the parameters as given in~\cite{ETG}, $V_R = 7.29$, $V_A = 3.18$,
$\mu_R = 614$~MeV and $\mu_A = 306$~MeV. From here on, we only
consider the S--wave part of this potential which can be obtained 
analytically. Although this potential
is quite simple, it captures essential features of the NN interaction,
in particular, it supports exactly one bound state at $E= -2.23$~MeV.
Since we are interested in
an effective theory with small momenta only, 
we set the cut--off $\Lambda = 400$~MeV (or smaller).
In Fig.~1, we compare  the original potential with the effective one.
The latter is defined via
\beq
{\cal V}_{\rm eff} = {\cal H}' - {\cal H}_0 \quad .
\eeq
In the range of the small momenta, the potentials are very similar.
However, one finds significant differences between the effective and
the original potential when the cut--off, above
which the nucleonic momenta are integrated out, is chosen very small,
$\Lambda \leq 200$~MeV. This is shown in Fig.~2. 

\medskip

\noindent {\bf 4.}
We now consider observables. Phase shifts can be derived from the
S--matrix, or equivalently, from a K--matrix approach. 
Symbolically, the relation between the S-- and the K--matrix
can be expressed as
\begin{equation}
{\rm S} = \frac{ 1 - i \pi q {\rm K}}{1 + i \pi q {\rm K}}~,
\end{equation} 
with $q$ the on--shell relative momentum of
the two nucleons corresponding to the laboratory energy being considered.
We work in the framework of the latter because the K--matrix is purely real. 
This is,  however, just a matter of convenience. 

The  low--energy phase shifts and  the bound--state  energy are
reproduced to a very high precision with the resulting effective 
potential acting only in the low momentum components. This is shown
for the S--wave phase in Fig.~3 for two values of the cut--off
$\Lambda = 200$ and $400$~MeV, respectively.
The phase shifts from the original potential are exactly reproduced in
the approach based on the effective potential, as long as one stays 
below the chosen cut--off. This is the reason why the solid and
dashed lines in the figure fall on top of each other.
The Lippmann--Schwinger equation in the effective approach involves
by construction only momenta below the cut--off and thus the bound
state can be calculated completely consistently. We find that
the bound state energy is also exactly reproduced. Furthermore, the
deuteron state evaluated in the effective theory is of course unitarily
transformed. Matrix--elements of an arbitrary operator
${\cal O}$ remain unchanged  under this unitary transformation,
\begin{equation}
\langle \Phi_D  | {\cal O} | \Phi_D  \rangle 
= \langle \Phi_D'  | {\cal O}' | \Phi_D'  \rangle~,
\end{equation}
with ${\cal O}' = U^\dagger {\cal O} U$, $ |\Phi_D'\rangle =
U^\dagger |\Phi_D \rangle$ and $ \Phi_D (p) = \langle p | \Phi_D
\rangle$ denotes the deuteron
wave function in momentum space. For illustration, we compare in 
Fig.~\ref{fwf}  the original and the unitarily transformed deuteron
momentum--space wave functions for $\Lambda = 200$~MeV. 
Note that due to the regularization,
 the ``spike'' close to the cut--off is infact a smooth function 
and does not introduce any singular derivatives. 
For a larger cut--off value, say 
$\Lambda =400$~MeV, the two curves fall onto of each other apart from a small
interval in the vicinity of the cut--off. We thus do not show that
case here. Note, however, that the unprojected deuteron 
wave function is still not negligible at momenta of about 800~MeV. As
an illustration, we consider the expectation value of the modulus of
the momentum operator in the S--wave deuteron, $\langle \Phi_D (p) |
\tilde{p} | \Phi_D (p) \rangle$, with $\tilde{p} =  |\vec{p} \, |$.
The results are listed in table~\ref{tab1}
for the two cut--off values $\Lambda =200, 400\,$MeV. Of course, for
the full MT--potential we do not need the regularization. However, to
illustrate its influence, we have also performed a calculation with a
MT--potential subject to the same regularization as used for the effective
potential (labelled ``regularized'' in the table). For the lower
cut--off, the few permille deviation between the exact result and the
one based on the regularized potential is simply due to the fact that 
we did not optimize the numerical solution of the 
integral equation~Eq.(\ref{eq:V}) to determine the operator $A$. If needed, one can
improve these numbers to agree to arbitrary precision (which is not of
relevance here).
\begin{table}[hbt]
\begin{center}
\begin{tabular}{|l|cc|}
    \hline\hline
    &  $\Lambda = 200\,$MeV  &  $\Lambda = 400\,$MeV  \\
    \hline 
    Exact Potential, not regularized      & 80.23 & 80.2308 \\
    Exact Potential, regularized          & 79.90 & 80.2303 \\
    Effective Potentail,  regularized  & 79.90 & 80.2304 \\
    \hline\hline
  \end{tabular}
\caption{Expectation value of $|\vec{p}\,|$ [MeV] in the S--wave deuteron
based on the exact and the effective MT--potential. For the exact
case, we show the results with and without regularization at the
singularity $|\vec{p}\,| = \Lambda$.}\label{tab1}
\end{center}
\end{table}

\medskip

\noindent {\bf 5.}
In summary, we have shown how to construct an effective low energy
theory for nucleons based on the method of unitary transformations.
For a simple S--wave potential,
we have shown that the theory projected onto  the  subspace of
momenta below a given cut--off reproduces exactly the features of the
original one. 
We hope that this study might be useful for derivation
of NN--forces based on chiral Lagrangians in the 
low--momentum regime. It should
also provide new insights into a consistent and convergent treatment 
of relativistic effects in few-- and many--nucleon systems.

\pagebreak



\pagebreak

\section*{Figures}

\vspace{5cm}

\begin{figure}[htb]
\centerline{\epsfig{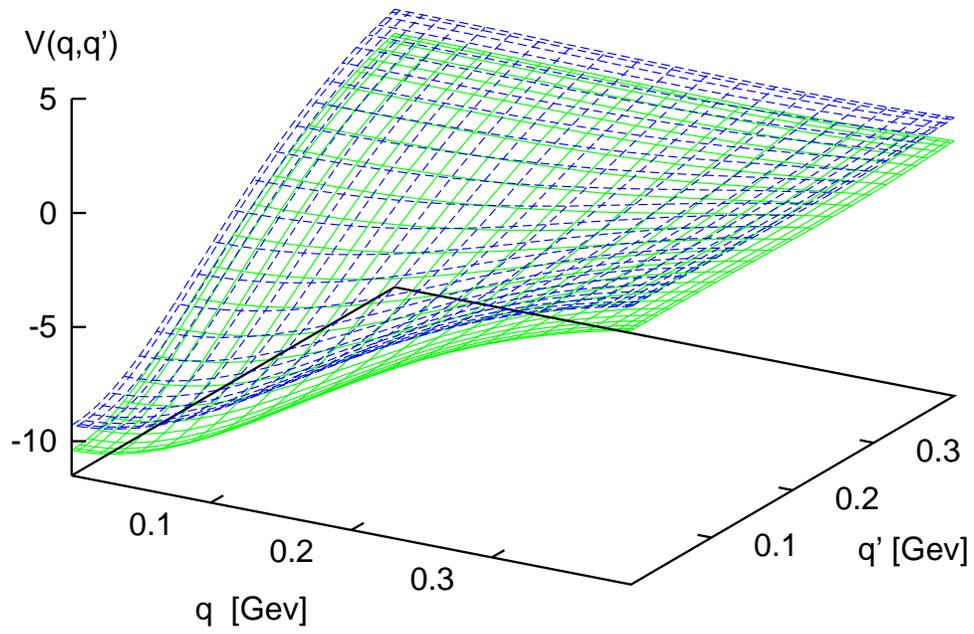}}

\vspace{3.3cm}

\caption[pot400]{\protect \small
Effective two--nucleon potential (green hatched area with solid lines)
in comparison with the original potential, Eq.(\ref{PotMT}) 
(blue hatched area with dashed lines), for momenta less than 400~MeV.}
\end{figure}

\begin{figure}[htb]
\centerline{\epsfig{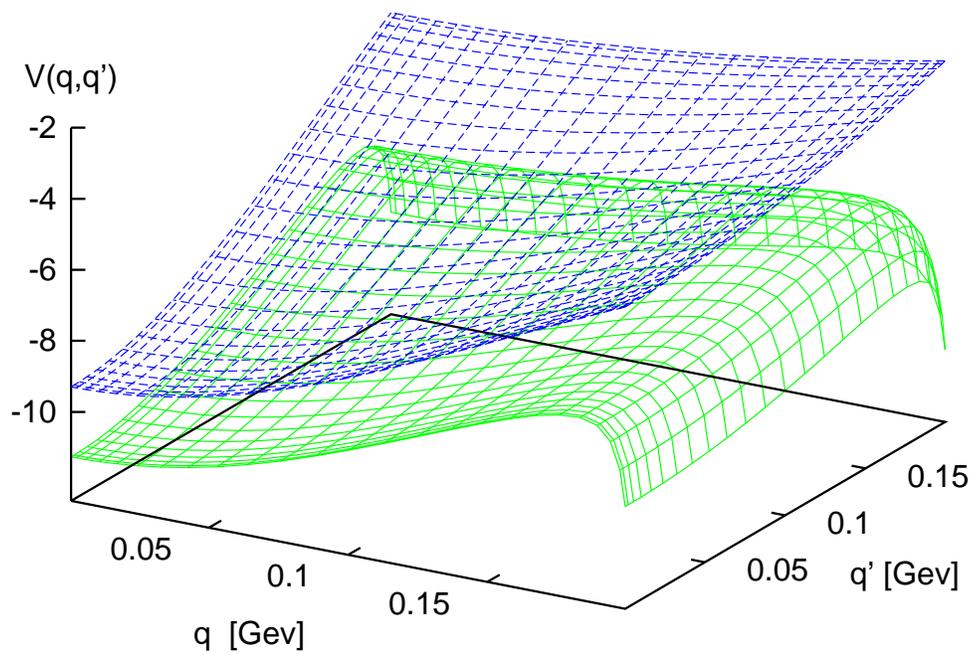}}

\vspace{3.3cm}

\caption[pot200]{\protect \small
Effective two--nucleon potential (green hatched area with solid lines)
in comparison with the original potential, Eq.(\ref{PotMT}) 
(blue hatched area with dashed lines), for momenta less than 200~MeV.}
\end{figure}

\begin{figure}[htb]
\centerline{\epsfig{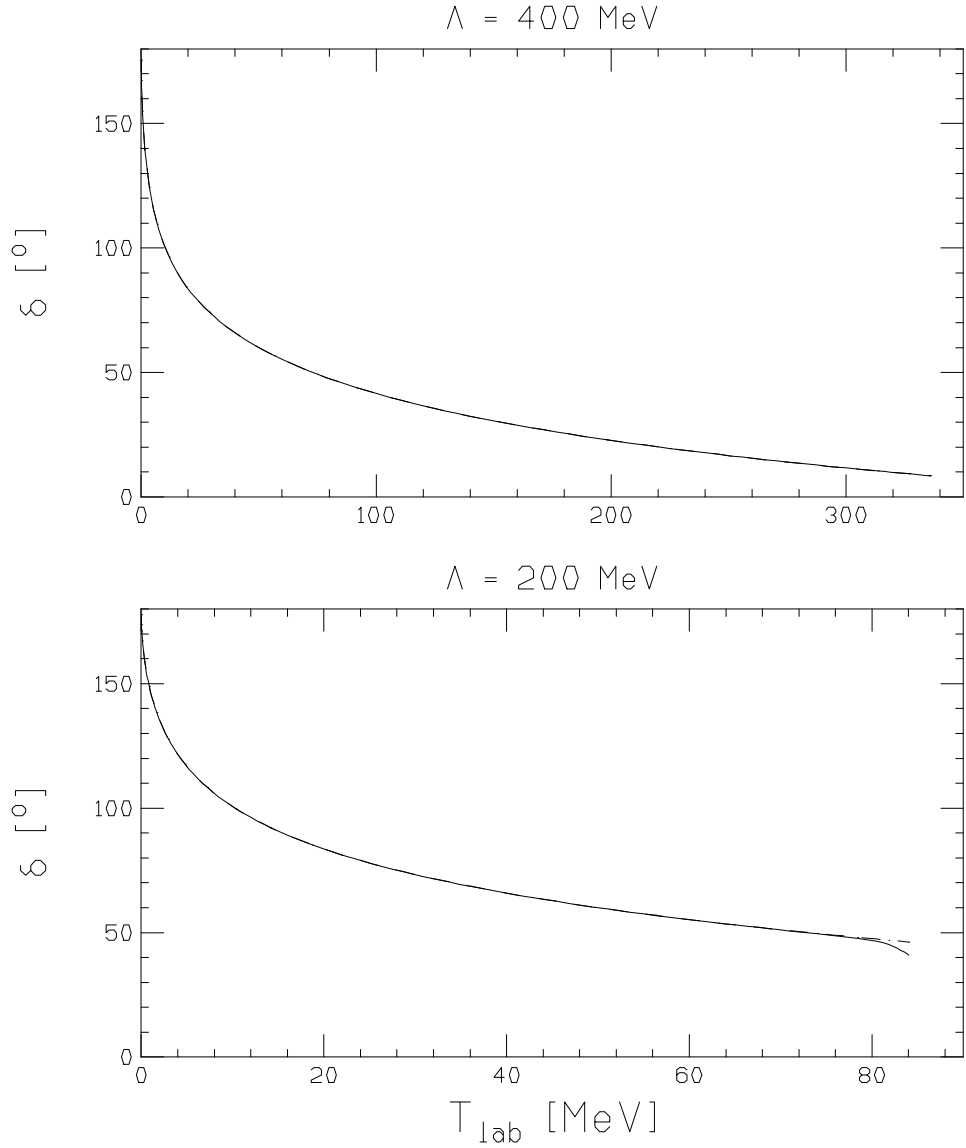}}

\vspace{3cm}

\caption[phases]{\protect \small
Phase shifts from the effective potential (solid lines) and the
original potential (dashed lines) as a function of the kinetic
energy in the lab frame. Upper (lower) panel: $\Lambda = 400\,
(200)$~MeV.}
\end{figure}

\begin{figure}[htb]
\centerline{\epsfig{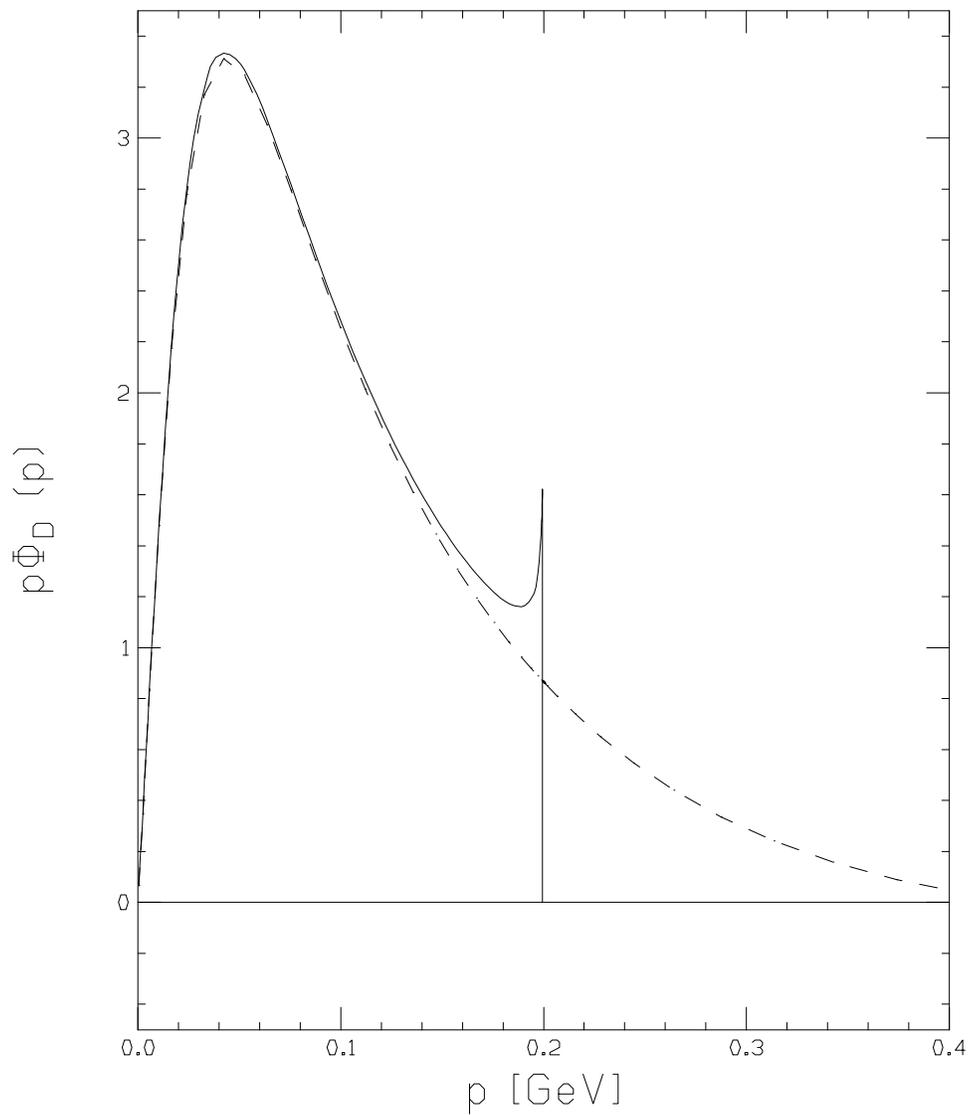}}

\vspace{3cm}

\caption[wf]{\protect \small \label{fwf}
Deuteron wave function $p \,\Phi_D (p)$ versus the momentum $p$
 from the effective potential (solid line) and the
original potential (dashed line) for $\Lambda = 200$~MeV.}
\end{figure}

\end{document}